\documentclass[super,pdflatex,sn-nature]{sn-jnl}%

\usepackage{graphicx}%
\usepackage{multirow}%
\usepackage{amsmath,amssymb,amsfonts}%
\usepackage{amsthm}%
\usepackage{mathrsfs}%
\usepackage[title]{appendix}%
\usepackage{xcolor}%
\usepackage{textcomp}%
\usepackage{manyfoot}%
\usepackage{booktabs}%
\usepackage{algorithm}%
\usepackage{algorithmicx}%
\usepackage{algpseudocode}%
\usepackage{listings}%
\usepackage[switch]{lineno} 

\theoremstyle{thmstyleone}%
\theoremstyle{thmstyletwo}%

\theoremstyle{thmstylethree}%

\newcommand\SI[2]{\ensuremath{#1\,\mathrm{#2}}}
\newcommand\um{\text{\textmu m}}
\newcommand\SiN{Si\ensuremath{_3}N\ensuremath{_4}\,}
\newcommand\SiO{SiO\ensuremath{_2}\,}

\makeatletter
\newcommand{\setbibnum}[1]{\setcounter{NAT@ctr}{#1}}
\makeatother

\begin{document}

\title[Article Title]{Super-resolution Optical Near-field EM for bio- and materials science}


\author[1,2]{\fnm{Ilia} \sur{Zykov}}
\equalcont{These authors contributed equally to this work.}

\author[3]{\fnm{Guido} \sur{Stam}}
\equalcont{These authors contributed equally to this work.}

\author[1,2]{\fnm{Hanieh} \sur{Jafarian}}
\equalcont{These authors contributed equally to this work.}

\author[3]{\fnm{Amin} \sur{Moradi}}
\equalcont{These authors contributed equally to this work.}

\author[3]{\fnm{Peter} \sur{Neu}}

\author[3,4]{\fnm{Rudolf} \sur{Tromp}}

\author[5]{\fnm{Mariana} \sur{Amaro}}

\author*[1,2]{\fnm{Thomas} \sur{Juffmann}}\email{thomas.juffmann@univie.ac.at}

\author*[3]{\fnm{Sense Jan} \spfx{van der} \sur{Molen}}\email{molen@physics.leidenuniv.nl}

\affil*[1]{\orgdiv{Faculty of Physics, VCQ}, \orgname{University of Vienna}, \orgaddress{\city{Vienna}, \country{Austria}}}

\affil*[2]{\orgdiv{Max Perutz Labs}, \orgname{University of Vienna}, \orgaddress{\city{Vienna}, \country{Austria}}}

\affil*[3]{\orgdiv{Huygens-Kamerlingh Onnes Laboratory, Leiden Institute of Physics}, \orgname{Leiden University}, \orgaddress{\city{Leiden}, \country{The Netherlands}}}

\affil[4]{\orgdiv{T.J. Watson Research Center}, \orgname{IBM}, \orgaddress{\city{New York},\country{USA}}}

\affil[5]{\orgdiv{J. Heyrovský Institute of Physical Chemistry}, \orgname{Czech Academy of Sciences}, \orgaddress{\city{Prague},\country{Czech Republic}}}


\abstract{Microscopy has been key to tremendous advances in science, technology, and medicine, revealing structure and dynamics across time and length scales. 
However, combining high spatial and temporal resolution in a non-invasive, label-free imaging technique remains a central challenge in microscopy. Here, we introduce Optical Near-field Electron Microscopy (ONEM), a method that converts optical near-field intensity patterns into photoelectron emission, enabling nanometer-scale imaging using low-energy electron microscopy.
ONEM achieves 31\,nm spatial and sub-second temporal resolution without exposing the sample to electrons, preserving structural and functional integrity. We demonstrate ONEM across three distinct domains: imaging polarization-dependent plasmon modes in metal nanostructures; visualizing live \textit{Escherichia coli} in liquid with orientation-resolved contrast in 3D; and capturing real-time electrodeposition of copper nanoclusters from solution. These results establish ONEM as a versatile platform for damage-free super-resolution imaging of interface dynamics in both vacuum and liquid, with broad implications for biology, electrochemistry, and nanophotonics.}

\keywords{Optical Near-field Electron Microscopy (ONEM), Label-free imaging, Non-invasive microscopy, Super-resolution, Plasmonics, Live-cell imaging, Electrochemical imaging}



\maketitle

\section{Introduction}\label{sec1}

Recently, optical microscopy based on elastic scattering has shown significant progress, enabling microsecond tracking~\cite{taylor_interferometric_2019}, mass photometry of single proteins~\cite{young_quantitative_2018}, and the imaging of cellular dynamics~\cite {kuppers_confocal_2023}. However, the spatial resolution is limited by the optical wavelength. The limit can be overcome using super-resolution techniques, mostly based on fluorescence, which can reach nanometric resolution. However, their temporal resolution is often limited by the fluorescence lifetime of the fluorophore, excitation scanning, or the reliance on temporal blinking behavior~\cite{schermelleh_super-resolution_2019, liu_super-resolution_2022}. Longer-term fluorescence imaging can furthermore be hampered by photo-toxicity~\cite{waldchen_light-induced_2015,laissue_assessing_2017}. 
While efficient labeling strategies have been developed for many biological applications, labeling is not always wanted or possible, \textit{e.g.}, in the harsh environments of electrochemistry experiments~\cite{hao_single-molecule_2020}. 
In electron microscopy, atomic-scale imaging is possible, both in dry~\cite{batson_sub-angstrom_2002} and liquid environments~\cite{williamson_dynamic_2003, liao_facet_2014, de_jonge_resolution_2019}. Cryo-electron microscopy and tomography enable the reconstruction of (static)
proteins~\cite{nogales_development_2016} and cellular architecture~\cite{mahamid_visualizing_2016}, respectively. However, electron-beam-induced damage generally prevents dynamic studies of sensitive samples~\cite{glaeser_how_2016}. Finally, scanning probe microscopy provides subatomic resolution~\cite{binnig_7_1983}. However, scanning the probe across the sample limits the imaging bandwidth and may perturb the sample~\cite{ando_high-speed_2018}. Similar concerns apply for near-field scanning optical microscopy~\cite{pohl_optical_1984}, where the resolution is limited by the probe's size, and data analysis is complicated by the probe's shape~\cite{yong_nano-spectroscopic_2018}.

Here, we experimentally demonstrate Optical Near-field Electron Microscopy (ONEM)~\cite{marchand_optical_2021}, a novel damage-free and label-free super-resolution imaging technique. 
The core idea of ONEM is to convert optical near-field intensities behind an illuminated object to photoelectrons, which are subsequently imaged using a low-energy electron microscope~\cite{tromp_new_2010, tromp_new_2013}. 
The spatial resolution thus obtainable is not given by the optical diffraction limit, but rather by the distance between the object and the photocathode~\cite{marchand_optical_2021}. Furthermore, optical near-fields (rather than far-fields) yields a fundamental increase of information that can be extracted about a sample~\cite {kienesberger_quantum_2024}. Importantly, ONEM has the advantage that an object is probed by photons only, and \textit{not} by (highly energetic) electrons. The sample never interacts with the photoelectrons, as these are emitted from a thin photocathode film, which is spatially separated from the sample itself.
This is the key to damage-free imaging, especially when visible light is used. 

We demonstrate ONEM with a spatial resolution better than \SI{31}{nm} or $\lambda/13$ for our illuminating wavelength of \SI{405}{nm}. We then show proof-of-principle experiments in plasmonics, live-cell imaging of \textit{E. coli} bacteria, and electrochemistry. We achieve sub-second temporal resolution, and show that the \textit{E. coli} bacteria are still alive after imaging them for hours. Our work shows the broad applicability of ONEM for in-situ interface studies.

\section{Results}\label{sec2}
\subsection*{Setup and Resolution}\label{subsec_setup}
Fig.~\ref{fig:setup}a shows a sketch of our ONEM system with the sample depicted in the center. The system comprises an optical excitation module and an aberration-corrected low-energy electron microscope (LEEM~\cite{tromp_new_2010, tromp_new_2013}) capable of imaging few-eV electrons with a lateral resolution of 1.5 nm~\cite{schramm_imaging_2013}.
A single-mode fiber delivers visible light at a wavelength of \SI{405}{nm} with a controlled polarization to the sample stage inside the ultra-high vacuum of the LEEM instrument. The light is collimated and illuminates the sample with an intensity of up to \SI{25}{mW/mm^2}. A weak pick-off beam is out-coupled and facilitates closed-loop polarization control.
The sample scatters and absorbs light, and the resulting near-field intensity pattern is converted into a spatially varying electron flux using a thin photocathode. We use both cesiated graphene~\cite{hu_cs_1986,gohler_transfer_2024} and cesiated chromium~\cite{chou_orbital-overlap_2012,fernandes_cauduro_tailoring_2019} photocathodes in this study. The cesiation is performed \textit{in situ} to prevent oxidation. 
The energy range of the emitted electrons is given by the difference between the photon energy (\SI{3}{eV} at \SI{405}{nm}) and the work function of the photocathode (\SI{2}{eV}~\cite{fernandes_cauduro_tailoring_2019}). The LEEM  instrument accelerates the photo-emitted electrons to \SI{10}{keV} and then images them onto a micro-channel plate detector. Further details about the setup are found in the Methods section.


First, we test the spatial resolution in ONEM by imaging a nanolithographically fabricated sample (Fig.~\ref{fig:setup}b). Alternating features of absorptive chromium, and transmissive fused silica are defined to create high optical contrast. 
Onto this sample, we apply a thin photocathode for visible light: a single layer of graphene with a sub-monolayer of cesium (see sketch in Fig.~\ref{fig:setup}b).
In Fig.~\ref{fig:setup}c, the fused silica appears as a bright line in the ONEM image, the chromium regions appear dark. A linecut through this image (indicated by the blue line), yields the intensity distribution plotted in Fig.~\ref{fig:setup}d. For a practical estimate of the resolution, we fit the plot with an error function $I=a\cdot \mathrm{erfc}\left(\frac{x-c}{\sigma\sqrt{2}}\right)+b$ (indicated by the red line), yielding $2\sigma= \SI{31}{nm}$. This provides an upper bound of the spatial resolution of ONEM, considering lithographic imperfections of the sample, the finite Cr thickness (\SI{30}{nm}), and the plasmonic response of the metal structure. Note that $\SI{31}{nm}$ corresponds to $\lambda/13$, \textit{i.e.}, well below the diffraction limit.


\subsection*{Plasmonics}\label{subsec_polarization}
Next, we present the capability of ONEM to observe light-matter interactions on the sub-wavelength scale. Specifically, we image localized surface plasmon resonances in \SI{100}{nm} cubic Ag nanoparticles on top of a $\SI{20}{nm}$ thin silicon nitride window (\SiN), coated with a \SI{3}{nm} caesiated chromium photocathode (see Fig.~\ref{fig:polarization}a). 
First, we ensure that we correctly identify Ag nanocubes in our ONEM images. We deposit \SI{500}{nm} \SiO beads as fiducials, which can easily be identified using both scanning electron microscopy (SEM) and ONEM at low magnification (Fig.~\ref{fig:polarization}b). We localize nearby Ag nanocubes using SEM (Fig.~\ref{fig:polarization}c) and image their characteristic polarization response with ONEM (Fig.~\ref{fig:polarization}d). We observe a globular shape for circularly polarized excitation and an elongated response for linearly polarized illumination.  

We then study the polarization response in more detail (Fig.~\ref{fig:polarization}e). For varying polarization of the illumination (top row), the orientation of the plasmonic mode measured in ONEM rotates (second row). The orientation of the plasmon can be identified with exposure times as low as \SI{1}{s} (see Extended Data Fig.~\ref{fig:polarization_1s}). 
The main features of these images agree qualitatively with finite-difference time-domain simulations using a 3D Electromagnetic Simulator \cite{noauthor_lumerical_nodate}  (third row, see the Methods section and Supplementary Note 1 for details). 
To obtain quantitative agreement of the contrast, we must convolve the simulations with a Gaussian of $2\sigma = \SI{35}{nm}$ (bottom row), comparable to the resolution estimated for the nanolithographic sample. 
We can identify two distinct features. The bright ellipse around each nanoparticle is related to constructive interference of the scattered field with the incoming excitation. Based on the average radius of the ellipse (400 nm) and the thickness of the window (23 nm), we can conclude that ONEM images collect information of light scattered into a solid angle of almost $2\pi$, representing a twofold improvement over conventional far-field high-numeric-aperture ($\mathrm{NA}=1.3$) imaging. More importantly, within the ellipse we observe the sub-diffraction features of the nanocube's plasmonic mode that do not propagate to the far field. A cross section of the $15^\circ$ polarization case is shown in Fig.~\ref{fig:polarization}f, together with a (broadened) simulation, all exhibiting three maxima. Interestingly, spherical nanoparticles show only two maxima, both in ONEM and simulations (see Fig.~\ref{fig:polarization}g). This demonstrates that ONEM can distinguish different shapes of sub-wavelength nanoparticles.


\subsection*{Live imaging of an \textit{E. coli} in solution}\label{subsec_liquid_bio}

To extend ONEM for dynamic investigations in solution, we designed a liquid cell compatible with the ultra-high vacuum system of the electron microscope which encloses a liquid volume of \SI{3}{mm^3} (see Fig.~\ref{fig:ecoli}a, the Methods section, and Extended Data Fig.~\ref{fig:liquid_cell}). 
To demonstrate liquid-cell ONEM for live-cell microscopy, we image the dynamics of \textit{E. coli} bacteria (RP437 strain) in Lysogeny Broth \cite{parkinson_isolation_1982} (see Supplementary Video 1). We observe both immobile and mobile bacteria, and can follow their trajectory over extended field-of-views using the motorized sample stage of the instrument. Importantly, the bacteria remain alive during and after ONEM imaging (see Supplementary Video 2), confirming the low invasiveness of this new super-resolution technique. Fig.~\ref{fig:ecoli}b shows the path of a mobile bacterium over the course of 22 seconds recorded at 3 Hz in a fixed field-of-view of about $50\times \SI{50}{\um}^2$. Selected frames from this trajectory are presented in Fig.~\ref{fig:ecoli}c-e, which show the bacterium in different 3D orientations at different times
. Comparing the images to simulations (Fig.~\ref{fig:ecoli}f-h) using the angular spectrum method in a multi-slice approach \cite{abdollahpour_four-dimensional_2011} yields the 3D orientation of the bacterium, in particular its orientation angle $\theta$ and distance $d$ to the photocathode (see Fig.~\ref{fig:ecoli}a). Simulation details can be found in the Methods section.  
Since the bacterium acts as a cylindrical lens, the projections become triangular for orientations that are neither horizontal nor vertical. 
So far, we have discussed straight bacteria only. However, we regularly observe bending of the bacteria, both in optical microscopy and ONEM, probably as a consequence of filamentation \cite{jaimes-lizcano_filamentous_2014}. We discuss these observations and related simulations in the Supplementary Note 2. 


\subsection*{Electrochemistry}\label{subsec_liquid_chem}
\begin{figure}[H]
	\centering
	\includegraphics[width=1.\textwidth]{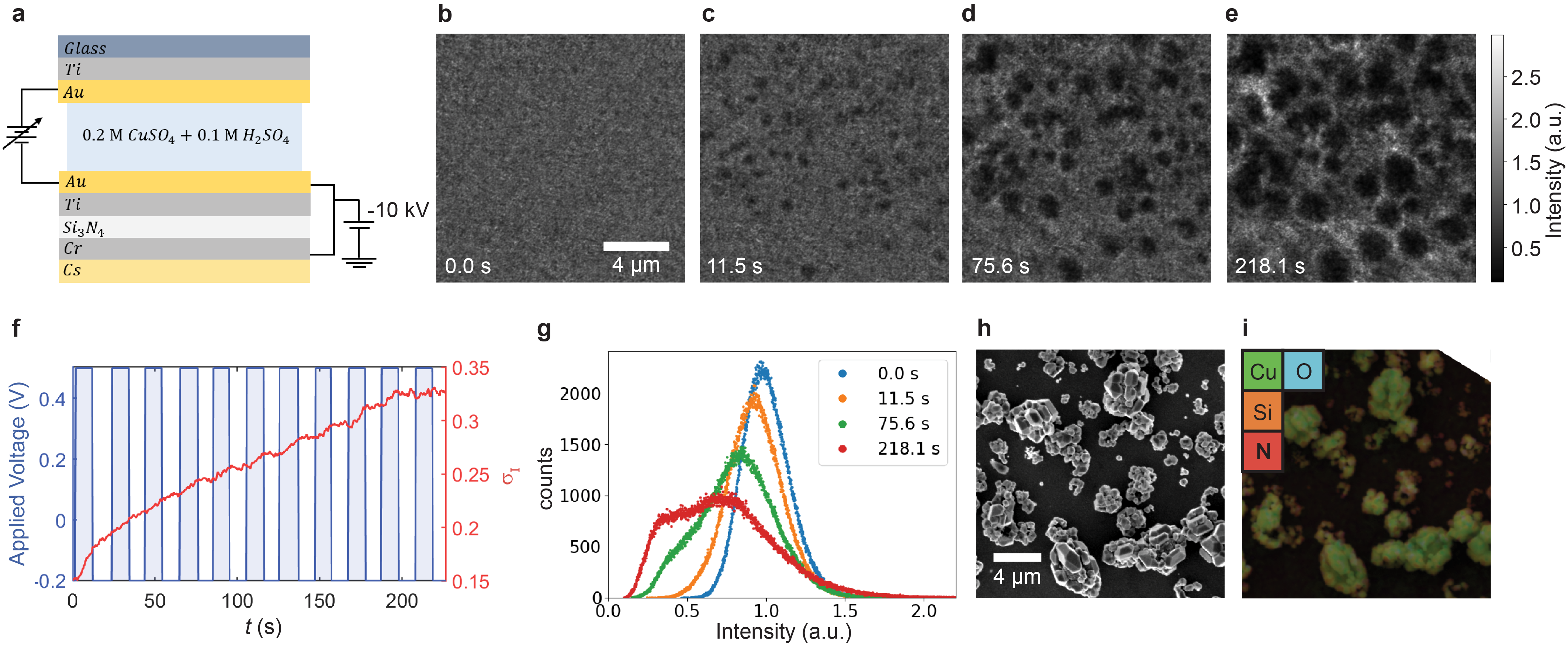}
	\caption{\textbf{Pulsed copper electrodeposition imaged with ONEM.}
        \textbf{a}, Schematic of ONEM electrochemical cell. The electrodes consist of 3\,nm Ti + 10\,nm Au on \SiN (WE), and on glass (CE). The WE is lifted to -10\, kV and a bias is applied to the CE vs. WE. \textbf{b-e}, Selected frames of an ONEM Cu electrodeposition experiment. \textbf{b}, initial image (t\,=\,0\,s). \textbf{c-e}, final frames of the first, fourth, and eleventh reduction cycle at t\,=\,11.5\,s, 75.6\,s, 218.1\,s, respectively. \textbf{f}, Standard deviation of the normalized intensity across the image, $\sigma_I$ (red), and potential applied to the setup (blue) vs. time. \textbf{g}, Intensity histograms corresponding to frames \textbf{b-e}. \textbf{h},  Post-mortem SEM image, with \textbf{i} corresponding EDS map of the same region of interest, confirming copper deposition.
    }    
    \label{fig:EC}
\end{figure}

To demonstrate ONEM's capabilities for real-time nanoscale imaging of electrochemistry, we record the nucleation and growth of copper nanoclusters on a gold electrode~\cite{williamson_dynamic_2003, radisic_morphology_2006, radisic_situ_2006}. 
Fig.~\ref{fig:EC}a shows a simple two-electrode liquid cell design compatible with ONEM. The liquid cell is filled with $\mathrm{0.2 \, M \, CuSO_4 + 0.1 \, M \, H_2SO_4}$ solution. The working electrode (WE) is deposited onto a silicon chip with a $\SI{20}{nm}$ thin \SiN window, and the counter electrode (CE) onto the glass window that seals the liquid cell. Both consist of a $\SI{10}{nm}$ thin Au layer on top of a $\SI{3}{nm}$ Ti adhesion layer. The WE is held at $\SI{-10}{kV}$ to enable electron optical imaging. A small potential is applied to the CE vs. the WE, facilitating the reaction $\mathrm{Cu^{2+} + 2e^- \rightleftarrows Cu}$.
We first determine the reduction and oxidation potentials of the system in a cyclic potential sweep experiment, demonstrating active control over copper deposition and stripping (see Supplementary Note 3).

Next, we visualize copper electrodeposition, using a sequence of voltage pulses (see blue curve in Fig.~\ref{fig:EC}f). Figs.~\ref{fig:EC}b-e show selected frames of an ONEM video of the WE (see Supplementary Video 4), recorded at a \SI{3}{Hz} frame rate, and post-processed as detailed in the Methods section. 
The red curve in Fig.~\ref{fig:EC}f shows the standard deviation of the normalized intensity across each video frame, $\sigma_I$, versus time. An increase in $\sigma_I$ indicates an inhomogeneous intensity distribution, \textit{e.g.}, due to copper nucleation and growth. This is what we observe during the first deposition cylce, after which, at $\SI{t=11}{s}$ (Fig.~\ref{fig:EC}c), we observe a set of copper islands with a typical radius of \SI{200}{nm}.
For the next three cycles, the increase in $\sigma_I$ continues, even between voltage pulses at the equilibrium potential, indicating that clusters continuously rearrange when the islands are still small. After the fourth deposition cycle, however, $\sigma_I$ shows step-like behavior, increasing only when the reduction potential is applied. Fig.~\ref{fig:EC}d shows a frame at this transition point. While their density remains constant, the islands show an increased size and a wider distribution of their widths and thicknesses.
Fig.~\ref{fig:EC}g shows the intensity histograms corresponding to the frames in Fig.~\ref{fig:EC}b-e. While the progressive shift to the left signifies a growth in the thickness of the islands, the depletion of the initial peak at high intensity signifies the lateral growth of the islands. Scattering and interference also lead to a small number of pixels that are brighter than those in the original distribution. 
A post-mortem SEM and Energy Dispersive X-Ray Spectroscopy (EDS) analysis on the same field of view is shown in Fig.~\ref{fig:EC}h and i, respectively. The SEM data shows how Cu islands have coarsened over time\cite{ross_liquid_2016}, which highlights the importance of observing electrochemical processes \textit{in situ} and in real-time. This is further emphasized by the EDS data, which shows that the copper particles have oxidized over time. Individual EDS maps are shown in the Extended Data Fig.~\ref{fig:EDX} and discussed in the Supplementary Note 4.

\section{Discussion}\label{sec12}

We have successfully implemented Optical Near-field Electron Microscopy (ONEM), a label-free super-resolution imaging technique. We obtain an upper bound for the spatial resolution of \SI{31}{nm}, \textit{i.e.} $1/13$ of the excitation wavelength of 405 nm. 
ONEM allows one to visualize the plasmonic response of metal nanoparticles to light of different polarizations. Plasmon orientation can be identified with exposure times as low as one second over a large field of view. While we demonstrated plasmon imaging of nanoparticles in vacuum, our liquid cell design allows for easy implementation in other environments. 
We further demonstrate the non-destructive imaging of living \textit{E. coli} bacteria inside a liquid cell. We imaged the bacteria for hours, acquiring dynamic data at sub-second temporal resolution, that encodes the 3D-orientation of the bacteria.
Finally, a proof-of-principle Cu deposition experiment demonstrates ONEM's ability to visualize dynamic, actively controlled electrochemical processes. 
In contrast to liquid-cell transmission electron microscopy, ONEM is not affected by electron-induced effects, such as radiolysis, that potentially alter the electrochemical behavior. 
Liquid-cell ONEM also offers a much larger liquid-cell volume ($10^3-10^6\,\times$), ensuring conditions closer to bulk experiments.

Future technological improvements will further increase the resolution, sensitivity, and speed of ONEM. Thinner support layers, ultimately graphene, can be used to separate the specimen and the photocathode, leading to higher contrast and better spatial resolution. More efficient photocathodes~\cite{gorlich_uber_1936, jensen_photoemission_2007, stam_growth_2024} will increase the photon-to-electron conversion efficiency by 2-3 orders of magnitude, which translates into higher frame rates and signal-to-noise ratios. These photocathodes also show high quantum efficiency when excited with green light, which will further reduce the invasiveness of ONEM imaging in biological applications.
Low energy direct electron detectors~\cite{wang_electron_2021} will increase detection quantum efficiency, and yield shot-noise limited measurement precision. Lastly, a dedicated three-electrode liquid cell with a potentiostat compatible with our ONEM electronics will enable quantitative electrochemistry experiments.

These advances will eventually enable the detection of proteins and protein assemblies~\cite{marchand_optical_2021}. One potential application lies in the study of molecular dynamics in lipid bilayers, which can be anchored close to the photocathode. 
The small point-spread function in ONEM  will enable the simultaneous detection and study of multiple scatterers within a diffraction-limited volume, and potentially also the study of conformational changes of molecular machines. 
ONEM is also a promising new technique for high-throughput applications such as connectomics: First, being non-invasive, ONEM allows for higher flux and fluence when imaging soft matter and biological material. Second, the photocathode in ONEM represents an area electron source, allowing for higher electron flux than from needle cathodes used in electron microscopy. 

Concluding, ONEM offers unique potential for fast, label-free, and non-invasive super-resolution imaging of interfaces. It opens the door to real-time in-situ studies of light-matter interactions, electrochemistry, (electro)catalysis, connectomics, and potentially cellular and molecular biology.





\begin{figure}[H]
	\centering
	\includegraphics[width=\textwidth]{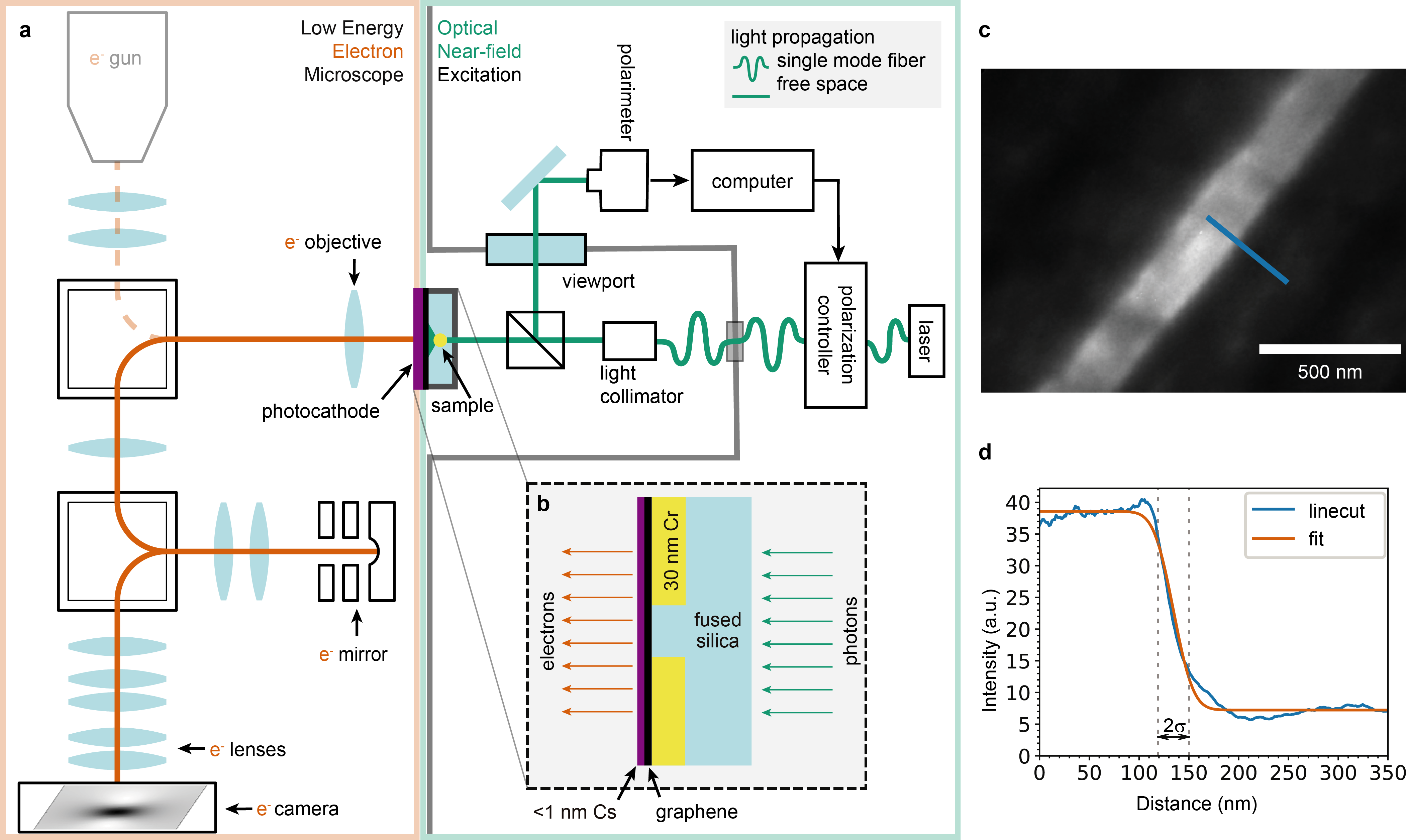} 
	\caption{\textbf{Optical Near-field Electron Microscopy (ONEM) setup and resolution.}
        \textbf{a}, Sketch: ONEM combines an optical illumination module (green) with electron detection in a low-energy electron microscope (LEEM, red). In ONEM, the electron gun of the LEEM system is off. Instead, the sample is illuminated with light from a single-mode fiber. The near-field intensities close to the sample are converted into a photoelectron flux using a photocathode. Closed-loop polarization control allows for setting any desired polarization of the illumination. 
        Electrons emitted by the photocathode are collected by the electron objective lens and imaged by the LEEM. \textbf{b}, ONEM on lithographically fabricated test sample: Photons pass through a fused silica substrate and interact with a geometrical pattern of chromium and fused silica. The resulting near-field intensities are converted to electrons in the cesiated graphene photocathode. \textbf{c}, ONEM image of a single \SI{250}{nm} wide line of the geometrical pattern at high magnification. \textbf{d}, An error function fit (red) to a (10 pixel wide) line cut (blue) of \textbf{c} yields an upper bound of the spatial resolution of $2\sigma=\SI{31}{nm}$.
    }       
	\label{fig:setup} 
\end{figure}

\begin{figure}[H]
    \centering
    \includegraphics[width=\textwidth]{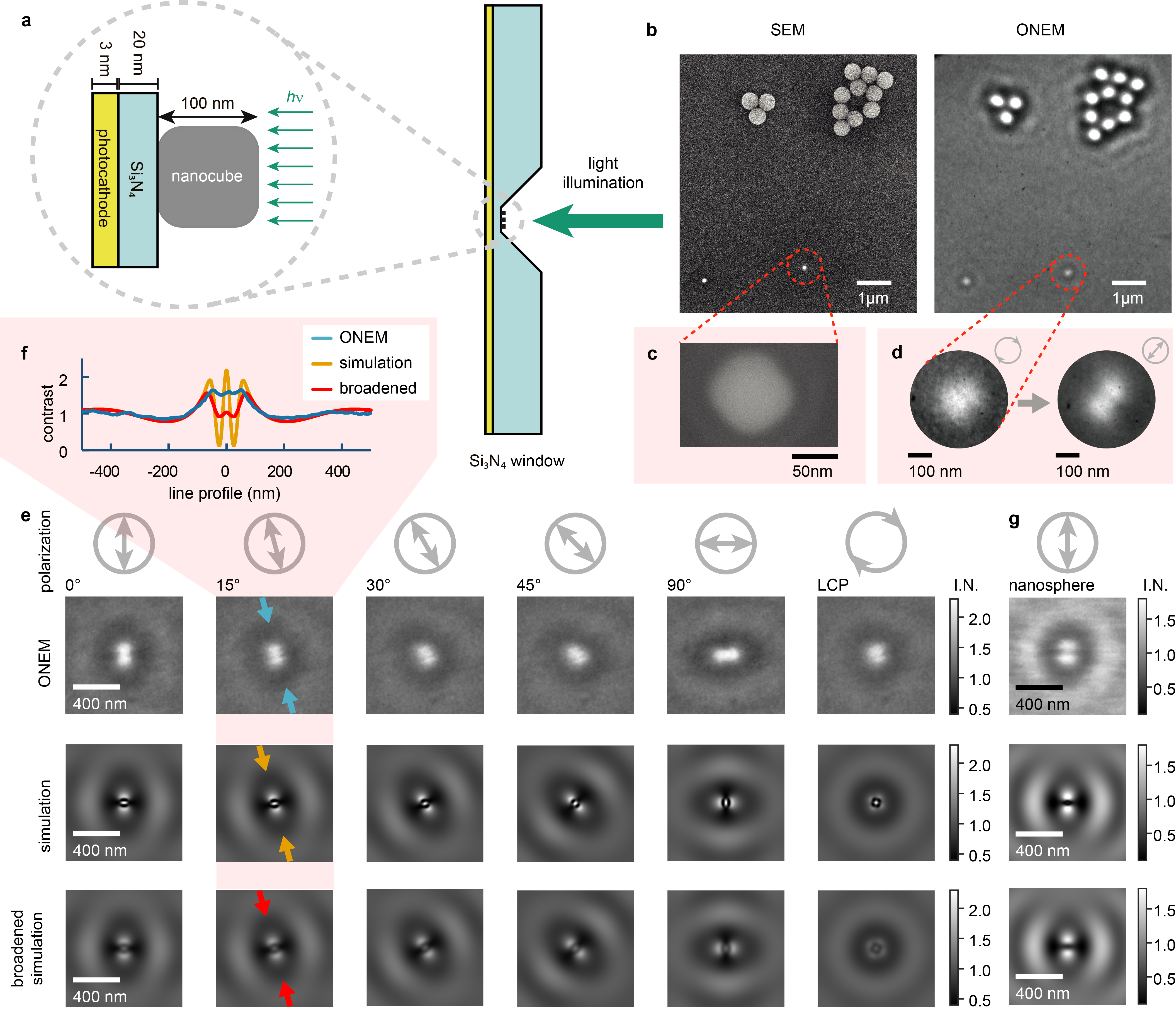}
    \caption{\textbf{Surface plasmon response imaged with ONEM.}
        \textbf{a}, Sample geometry. \SI{100}{nm} Ag nanocubes on top of \SI{20}{nm} \SiN are illuminated with \SI{405}{nm} collimated light. A \SI{3}{nm} layer of caesiated chromium acts as a photocathode. \textbf{b}, Sample area imaged by both SEM (left), and ONEM using circularly polarized illumination (right). \textbf{c}, Zoom: nanocube in SEM. \textbf{d}, Magnified ONEM images of nanocube using circular (left) and linear (right) polarization. \textbf{e}, Polarization response of nanocube. Top row: (linear) polarization angles chosen, except: LCP - left circular polarized. Second row: ONEM images, showing the intensities normalized to the background (I.N). Third row: simulated intensities in the photocathode. Fourth row: third row convoluted with a Gaussian with $2\sigma$ = \SI{35}{nm}. 
        \textbf{f}, Line profile of images in second column of \textbf{e}, taken between arrows, comparing ONEM and (broadened) simulation. Note the hot spot in the center. \textbf{g}, ONEM image and (broadened) simulation of a nanosphere illuminated with linearly polarized light. Note the minimum in the center.}
\label{fig:polarization} 
\end{figure}

\begin{figure}[H]
    \centering
    \includegraphics[width=\textwidth]{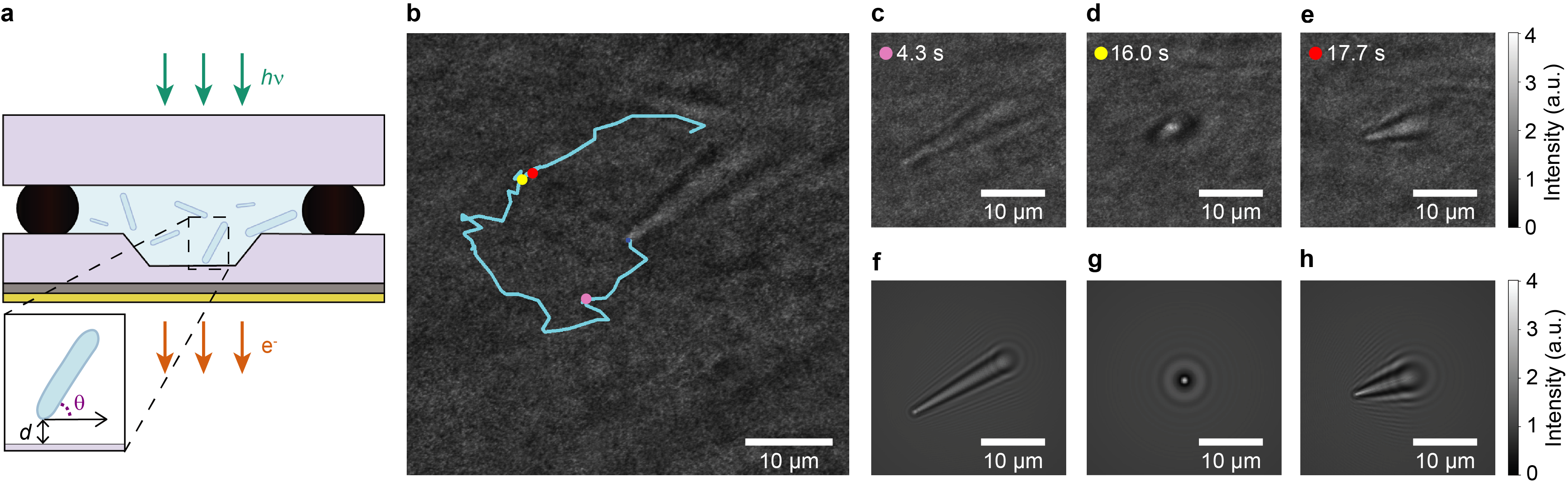}
    \caption {\textbf{Live \textit{E. coli} bacteria imaged with ONEM.} \textbf{a}, Schematic of the ONEM liquid cell.  \textit{E. coli} bacteria (blue capsules, not to scale) are confined between a glass window and a \SiN chip, separated by an O-ring.  The closest distance of the bacterium to the photocathode, consisting of Cr (dark gray) and Cs (yellow), is indicated by $d$, while its orientation relative to the photocathode plane is given by the angle $ \theta$.
    \textbf{b}, the bacterial trajectory (blue) over a 22-second recording, starting at \SI{t=0.3}{s} when the bacterium enters the field-of-view. \textbf{c, d, e}, selected frames. \textbf{f, g, h}, simulations for a \SI{20}{\um} long and \SI{1}{\um} wide bacterium \textbf{f} with $d= \SI{1}{\um}, \theta = 30^\circ$, \textbf{g} with  
    $d= \SI{3.15}{\um}, \theta = 90^\circ$, and \textbf{h} with $d= \SI{0.5}{\um}, \theta = 60^\circ$.} 
\label{fig:ecoli}
\end{figure}  

\section{Methods}\label{methods}

\subsection*{Optical module design}
The ONEM microscope uses laser light at a wavelength of \SI{405}{nm} (Vortran Stradus laser diode, \#405-100) to probe the sample. The light is in-coupled into the single-mode fiber (Thorlabs, 400 - 680 nm, \#S405-XP) that leads into the vacuum chamber via a feedthrough. Inside the vacuum chamber, the light is out-coupled from the fiber and collimated. The collimator consists of a single lens (Edmund Optics' \SI{5.0}{mm} Diameter, \SI{10}{mm} focal length, VIS-NIR coated, plano-convex lens, \#47-858) positioned within the aluminum body of the collimator assembly.
The fiber is terminated by a stainless steel ferrule with a set screw fixing the distance from the fiber tip to the lens. The collimator is followed by a non-polarizing beamsplitter (Thorlabs's UV-grade fused silica plate beamsplitter with 400-700 nm coating, 1/2 inch diameter, \SI{3}{mm} thick, \#BSN04) that transmits about 90\% of the light intensity towards the sample, and reflects the rest towards a polarimeter (Thorlabs, \#PAX1000VIS/M). The transmitted light reaches the sample after free propagation as a Gaussian beam. It has a full-width half-maximum of \SI{0.2}{mm} and central intensity of up to \SI{25}{mW/mm^2} at the sample plane and a convergence angle of less than \SI{20}{mrad}. The reflected light reaches the polarimeter (Thorlabs's Polarimeter, \#PAX1000VIS/M) outside of the vacuum chamber via a viewport (low stress-induced birefringent Zerodur glass) and broadband dielectric mirrors (Thorlabs' fused silica mirrors, \#BB07-E02 and \#BB2-E02). The collimator and beamsplitter are fixed in a 3D-printed titanium body with spring-loaded bayonet connectors for self-alignment to the sample holder and an interface for grabbing it with a transfer arm. This allows the collimator to be removed during sample transfers and stored within the sample measurement chamber. The choice of materials and components was dictated by the \SI{0.2}{N} limitation in the load that can be applied to the 5-axis sample stage, the requirement to have no magnetic metals in the assembly, compatibility with ultra-high vacuum, and sample chamber size constraints. See Extended Data Fig.~\ref{fig:setup_solidworks} for the detailed sketch of the illumination part of the setup.

\subsection*{SEM and EDS}
The scanning electron microscopy (SEM) data in Fig.~\ref{fig:polarization} are collected using an Apreo scanning electron microscope from Thermo Fisher. The SEM and energy-dispersive X-ray spectroscopy (EDS) data in Fig.~\ref{fig:EC} and Extended Data Fig.~\ref{fig:EDX} are collected using a Zeiss Supra 55 VP. 

\subsection*{Polarization control}
The polarization feedback loop consists of a fiber-based polarization controller (OZ Optics' Electrically Driven Polarization Controller-Scrambler) and a polarimeter. The polarimeter reads the polarization of the light that is out-coupled by a 90-10 beamsplitter located just before the sample. Considering the birefringence of the beamsplitter, the viewport, and the mirrors, the polarization of the illumination can be calculated. Any desired state of polarization of the illumination beam is reached via a closed loop based on the measured polarization.
It has a relative error in Stokes parameters typically below 2\%. Since the polarization state drifts at a time scale much longer than the duration of the experiments, we use the polarization control only to set a new state of polarization.

\subsection*{Photocathode growth}
The photocathodes used in this paper comprise a thin support layer (graphene or chromium) with a sub-monolayer of Cs. The adsorption of Cs decreases the work function of many materials down to $1.6$-$\SI{2.0}{eV}$~\cite{hu_cs_1986,gohler_transfer_2024,chou_orbital-overlap_2012,larue_work_2008,fernandes_cauduro_tailoring_2019}, enabling the creation of photoelectrons with visible light.
The Cs is thermally evaporated onto the substrate inside the ultra-high vacuum chamber of the ONEM setup. This is done using an evaporator equipped with a small dispenser containing a cesium molybdate salt from SAES getters.
\subsection*{Liquid cell ONEM}
To image liquid-phase systems, we constructed an ultra-high vacuum-compatible liquid cell (LC). Below, we describe the design of the ONEM LC and its adaptation for electrochemistry experiments. 

\subsubsection*{Liquid cell design} \label{sup_LC_design}
The ONEM liquid cell consists of an inner and an outer shell (see Extended Data Fig.~\ref{fig:liquid_cell}a). The inner shell encloses the liquid. It is composed of a silicon chip with a \SI{20}{nm} \SiN window (Silson, M1000085), a viton O-ring (Arcus Dichtelemente GmbH, catalog no. 10838), and a~\SI{1}{mm} thick B270 window (Edmund Optics, 64-083).
The O-ring is$~\SI{0.64}{mm}$ thick and has an inner diameter of$~\SI{3.25}{mm}$, which is compressed to approximately half its thickness to ensure a stable seal against ultra-high vacuum. Through drop casting, a liquid volume of about$~\SI{3}{mm^3}$ is encapsulated. This volume can be adjusted by selecting a different size of the Viton O-ring. 
The flat side of the \SiN chip is coated with the photocathode layer and faces the electron optics. 
The outer shell provides structural support and compresses the O-ring, forming a static seal secured by a threaded fastening. It is machined out of aluminum, providing a low mass and minimal magnetic susceptibility. Extended Data Fig.~\ref{fig:liquid_cell}b shows the liquid cell alongside a 1 euro coin to provide a sense of scale and perspective. A central hole in the outer shell design ensures the passage of the illumination light to the glass window, and subsequently through the liquid to the \SiN membrane.

\subsubsection*{ONEM Liquid Cell for electrochemistry} 
For electrochemical experiments, we designed a liquid cell that features two electrodes. For the working electrode, we use custom-coated \SiN chips from Silson (\SI{3}{nm} Ti + \SI{10}{nm} Au). For the counter electrode, we coat the glass window with (\SI{3}{nm} Ti + \SI{10}{nm} Au) using a Mantis Hex deposition system. In both cases, Ti acts as an adhesion layer between gold and the underlying substrate (glass or \SiN). A reliable electrical contact to these electrodes is achieved using UHV-compatible Kapton-insulated wire with an outer diameter of \SI{0.71}{mm} (Kurt J. Lesker, FTAK06010) with fixed pogo pins (MULTICOMP, P50-H-120-G) at its ends. They are secured using UHV-compatible heat-shrink tubing (Allectra, 316-SHRINK-20), which provides both insulation and mechanical stability. This can be seen in Extended Data Fig.~\ref{fig:liquid_cell}c. The Kapton wire connects to the electrical contact points of the LEEM sample holder.

\subsection*{Sample preparation}
Here, we describe the sample preparation procedures used for the samples, in order of their appearance in the figures of the manuscript.

\subsubsection*{Nanolithography Sample (Fig.~\ref{fig:setup})}

After cleaning pre-cut $6\times \SI{6}{mm}$ fused silica substrates with acetone and then isopropanol, a layer of negative resist (ARN 7500) is applied by spin coating (1 minute at 4000 rpm with a baking step of 1 minute at \SI{85}{^{\circ} C}). Next, a standard geometrical pattern is written with e-beam lithography in an EBPG Raith-100 instrument. The sample is then developed in AR 30047 for 30 seconds. This step is repeated if the negative resist is not yet fully developed as visible by inspection with an optical microscope. To remove residual undeveloped resist in small corners of the sample, the sample is descummed by oxygen plasma for 4 minutes in a plasma cleaner (Oxford PlasmaLab 90+). Then, \SI{30}{nm} of silica is etched away with an O$_2$/Ar/CHF$_3$ etchant recipe (Extended Data Fig.~\ref{fig:lithography_prep}b) and filled up again with Cr from a thermal evaporator, shown in Extended Data Fig.~\ref{fig:lithography_prep}c.
Subsequently, the sample is placed into acetone at \SI{45}{^{\circ} C} for 24 hours to lift off the developed negative resist. Graphene is transferred onto the sample using a recipe developed by the supplier (Graphenea), and lastly, a sub-monolayer of Cs is deposited. This last step is performed in ultra-high vacuum inside the ONEM microscope. Extended Data Fig.~\ref{fig:lithography_prep}d shows a sketch of the final sample geometry.

Extended Data Fig.~\ref{fig:lithography_intermediate_short} depicts the sample imaged with ONEM (a) and with an optical microscope (b). The small squares in the structure have sides that are \SI{10}{\um} long and \SI{10}{\um} wide. Within each square, lines have been written, each \SI{2}{\um} apart and roughly \SI{250}{nm} wide. The resolution analysis in the paper is done on one of the thin lines. In the optical image, coarse errors in the sample fabrication are visible. In the ONEM image, we observe that grain boundaries and graphene double layers modulate the photoemission from the transferred graphene photocathode.

\subsubsection*{Plasmonics sample (Fig.~\ref{fig:polarization})}
Silicon chips (Silson, \SI{0.2}{mm} thick, 5 mm wide) with \SiN windows (thickness \SI{20}{nm} for Fig.~\ref{fig:polarization}b-e and \SI{30}{nm} for Fig.~\ref{fig:polarization}g, membrane size 1x1 mm) were coated with \SI{3}{nm} of chromium on the front (flat) side.
Silver nanocubes and \SiO beads were deposited in a series of nebulizer (Beurer IH57 mesh inhalator) evaporation steps onto the back side of the chip as shown in~Fig.~\ref{fig:polarization}a. Silver nanocubes were deposited from 6 ml of \SI{20}{\text{\textmu} g/ml} silver nanocube solution freshly dissolved in water from a stock of \SI{2}{mg/ml} of Nanocomposix's \SI{100}{nm} silver nanocubes (SCPH100). The \SiO beads were deposited from \SI{2}{ml} of \SI{64}{\text{\textmu} g/ml} \SiO solution freshly dissolved in isopropanol from a stock of 5\% Sigma-Aldrich's \SI{500}{nm} silica nanoparticles (913618).

\subsubsection*{\textit{E. coli} (Fig.~\ref{fig:ecoli})}
The \textit{E. coli} in Lysogeny Broth (LB) is received from Ariane Briegel's group at Leiden University, 
at an optical density of $\mathrm{OD}_{600}=1$. The sample is stored at \SI{4}{^{\circ}C} for 4 days prior to the measurement. Before the experiment, \SI{40}{\text{\textmu} L} of the sample is diluted in \SI{1}{mL} LB. Once the liquid cell is assembled, it is loaded into the ONEM setup. Pumping and microscope alignment are completed within 2 hours, after which we image the sample for about \SI{5}{h} at about \SI{10}{\%} of the full laser power. Subsequently, we observe the sample within the liquid cell using an optical microscope (Supplementary Video~2).

\subsection*{Image processing}
In this section, we describe the image post-processing carried out on the ONEM data. 

\subsubsection*{Lithography sample (Fig.~\ref{fig:setup})}
The ONEM images (20 images, 4-second exposure each) are detector corrected and drift corrected according to de Jong \textit{et al.} \cite{de_jong_quantitative_2020} and are averaged to obtain the single image shown in Fig.~\ref{fig:setup}c.

\subsubsection*{Plasmonics data (Fig.~\ref{fig:polarization})}
To obtain the images shown in Fig.~\ref{fig:polarization} we average 32 frames after drift-correction and normalization to the background. Each frame has an exposure time of 1 second. A background image is obtained by laterally translating the sample such that there are no nanoparticles in the field of view.

\subsubsection*{\textit{E. coli} data (Fig.~\ref{fig:ecoli})}
The image post-processing of this data set consists of three steps. First, Fourier filtering is applied to all frames to remove the standing-wave pattern caused by interference with optical reflections from the slanted side walls of the \SiN chip (see the magnified image in Extended Data Fig.~\ref{fig:liquid_cell}a). Second, the median intensity across time is calculated for each pixel and used as the background of the entire series. Then, each frame is divided by this background. Additionally, the trajectory in Fig.~\ref{fig:ecoli}b in the main text is obtained by manually identifying the front of the bacterium in each frame.
\subsubsection*{Pulsed electrodeposition data (Fig.~\ref{fig:EC})}
 First, all data is Fourier filtered to suppress interferences due to reflections from the slanted walls of the silicon chip. The frames are then normalized by the background image. The background image is the average of the first 10 frames at \SI{-0.2}{V} potential, where no electrochemical reaction took place due to the equilibrium. Next, the images are smoothed using a three-frame temporal sliding window. A median filter (kernel size = 3 pixels) and a Gaussian blur filter (kernel size = 9 pixels) are applied to further reduce noise while preserving intensity information.
\subsubsection*{Cyclic potential sweep experiment (Extended Data Fig.~\ref{fig:cv})}
The data processing applied to this measurement (Extended Data Fig.~\ref{fig:cv}) is the same as described above for the pulsed electrodeposition data. However, the background is calculated using the average of the last 10 frames, in which the WE is clean of Cu. These 10 frames are not included in the data presented in Extended Data Fig.~\ref{fig:cv}. 

\subsection*{Simulations}
In this section, we explain the simulation procedures used in the manuscript.

\subsubsection*{Plasmonics sample simulation (Fig.~\ref{fig:polarization})}
The finite-difference time-domain (FDTD) simulations were performed in a \SI{4}{\um} by \SI{4}{\um} by \SI{2}{\um} volume, with the last being perpendicular to the photocathode. Periodic boundary conditions were used for the boundaries perpendicular to the photocathode. Perfectly matched layer boundary conditions were used for the boundaries parallel to the photocathode. To simulate the nanocube, the shape of a rectangular cuboid with rounded edges was used, as observed in the corresponding SEM images. This yielded a side length of \SI{91}{nm} and a radius of curvature of \SI{27}{nm}. The silicon nitride membrane and photocathode were simulated as homogeneous layers of corresponding thickness.

\subsubsection*{\textit{E. coli} simulation (Fig.~\ref{fig:ecoli})}
The simulations for the bacterium were implemented in MATLAB. The bacterium is modeled as a homogeneous capsule, with a length of \SI{20}{\um} and a radius of \SI{0.5}{\um}, having a refractive index of $n_B = 1.38$. It is immersed in the LB medium with a refractive index of 1.33. 
Since $n_B$ is real, and since the size of the bacterium is larger than the illumination wavelength, image contrast arises from interference of the incoming field with the propagating scattered fields. For modeling, we assume plane-wave illumination with a Gaussian intensity distribution (beam waist $\omega_\circ= \SI{80}{\um}$), and a wave vector vertical to the screen. 




\section{Conclusion}\label{sec13}

\backmatter

\bmhead{Acknowledgements}
We thank all members of the ONEM collaboration for useful discussions. We thank Ariane Briegel and  Sidi Mabrouk at Leiden University for providing the E. coli.  We thank Esra Şimşek and Noud van Halteren for their assistance in imaging during the experiments.
We thank Philipp Redfern for assistance in the installation of the polarization control module.
We thank Bram Vanhecke for assistance with the simulation of the parabolic bacterium.
We thank Constanza Cendon Contreras for assisting in making the electrical contacts for electrochemical ONEM LC.
We thank the Electron Microscopy Facility of the Vienna BioCenter Core Facilities GmbH (VBCF), member of the Vienna BioCenter (VBC), Austria, for providing access to the transmission electron microscope, used for sample preparation.

\bmhead{Funding}
This project has received funding from the European Union’s Horizon 2020 research and innovation programme
under grant agreement No 101017902. Additional funding came from the Dutch 'Zwaartekracht' program 
ANION.

\bmhead{Author contribution}
These authors contributed equally: Ilia Zykov, Guido Stam, Hanieh Jafarian, Amin Moradi.

Conceptualization: T.J., S.J.v.d.M., R.T.

Methodology: I.Z., G.S., H.J., A.M., P.N., T.J., S.J.v.d.M., R.T.

Experiments: I.Z., G.S., H.J., A.M., P.N.

Visualization: I.Z., G.S., H.J.

FTDT Simulations: I.Z.

Angular Spectrum Representation Simulations: H.J.

Development of Optical Illumination and Polarization Control Module: I.Z., T.J.

Photocathode Development: G.S., A.M., P.N., S.J.v.d.M., R.T.

Liquid Cell Design: H.J., T.J.

Electrochemical Cell Design: H.J.

Lithographic Sample Preparation and Characterization: G.S., A.M., P.N.

Plasmonic Sample Preparation and Characterization: I.Z. 

Biological Sample Preparation and Characterization: H.J.

Electrochemical Sample Preparation and Characterization: H.J.

Data analysis: All authors

Writing – Original Draft: I.Z., G.S., H.J., P.N., R.T., S.J.v.d.M., T.J.

Writing – Review and Editing: All authors

Supervision: T.J., S.J.v.d.M., R.T.

Funding Acquisition: M.A., S.J.v.d.M., T.J.

\bmhead{Conflict of interest/Competing interests}
There are no competing interests to declare.

\bmhead{Materials \& Correspondence}
Correspondence to Thomas Juffmann and Sense Jan van der Molen.

\bmhead{Supplementary information}

The article contains Supplementary Notes 1 to 4 and Supplementary Videos 1 to 5.


\renewcommand{\figurename}{Extended Data Fig.}
\setcounter{figure}{0}

\begin{figure}[H]
	\centering
	\includegraphics[width=0.5\textwidth]{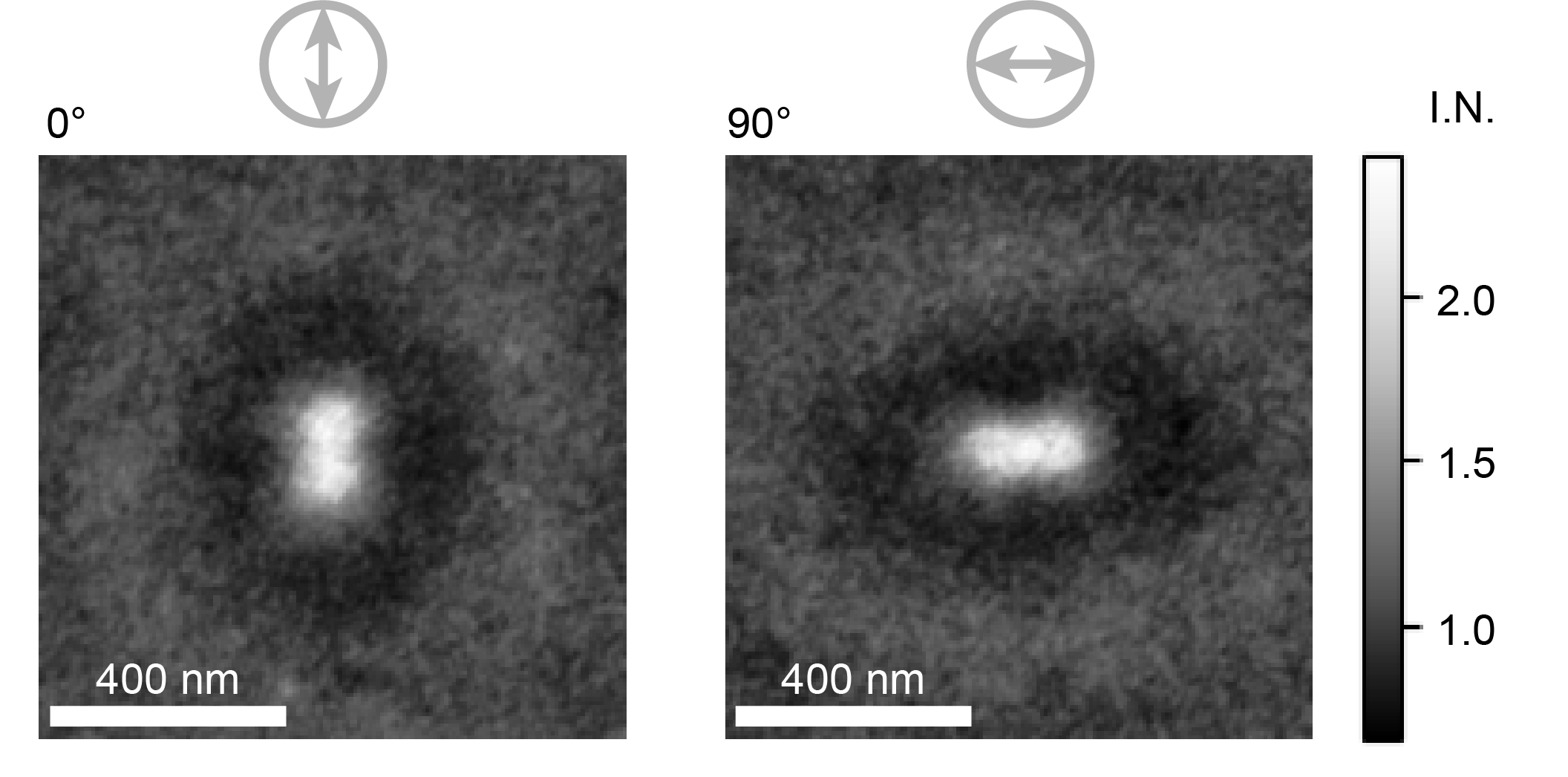}
	\caption{\textbf{Demonstration of ONEM capability to distinguish the orientation of plasmons within 1~second acquisition time.} ONEM image of a \SI{100}{nm} Ag nanocube illuminated with two perpendicular linear polarizations. Exposure times of 1 second are sufficient to determine the orientation of the plasmon. I.N. - intensity normalized to the background.
    }
	\label{fig:polarization_1s} 
\end{figure}

\begin{figure}[H]
\centering
\includegraphics[width=\textwidth]{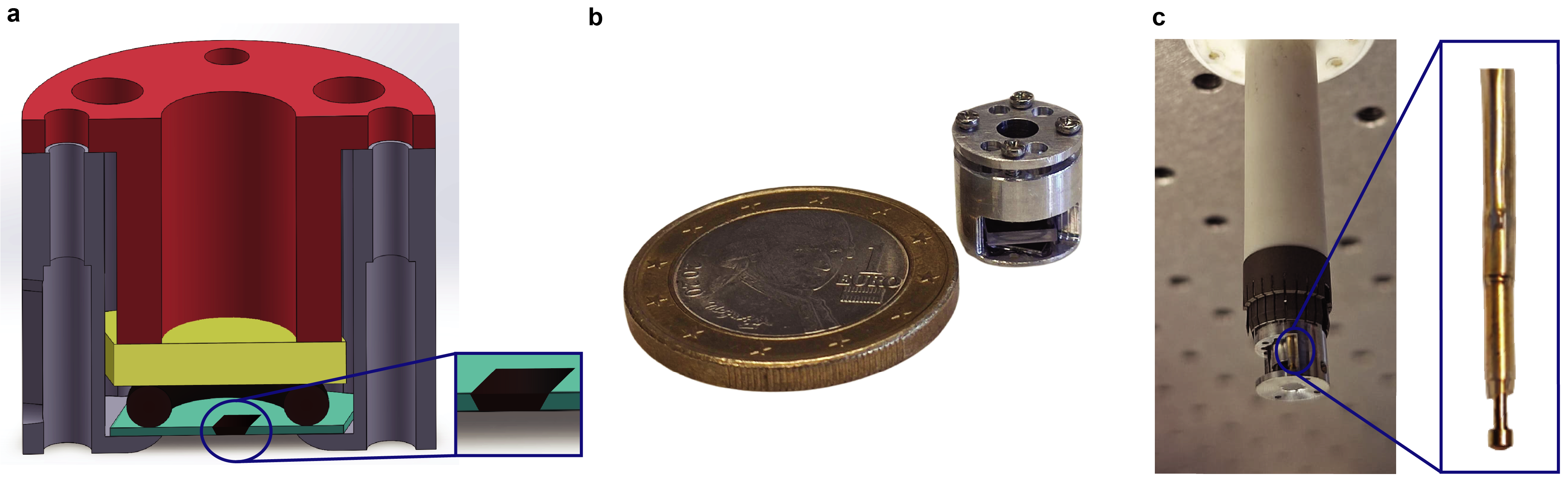}
\caption{\textbf{ONEM liquid cell.} \textbf{a}, Cross-cut of the ONEM liquid cell, showcasing both the outer shell (in red and gray) and the inner shell (from bottom to top \SiN chip, O-ring, and glass window). The inset shows the \SiN window, with slanted Si walls next to it. Light reflections from these walls lead to interference patterns in our ONEM images, which are removed using Fourier filtering. \textbf{b}, The assembled liquid cell is displayed next to a 1 Euro coin. \textbf{c}, The liquid cell is placed at the tip of the LEEM sample holder. The magnified image shows a pogo pin pressing against the coated \SiN chip. It ensures a secure electrical contact required for electrochemical experiments. } 
\label{fig:liquid_cell}
\end{figure}

\begin{figure}[H]
\centering
\includegraphics[width=\textwidth]{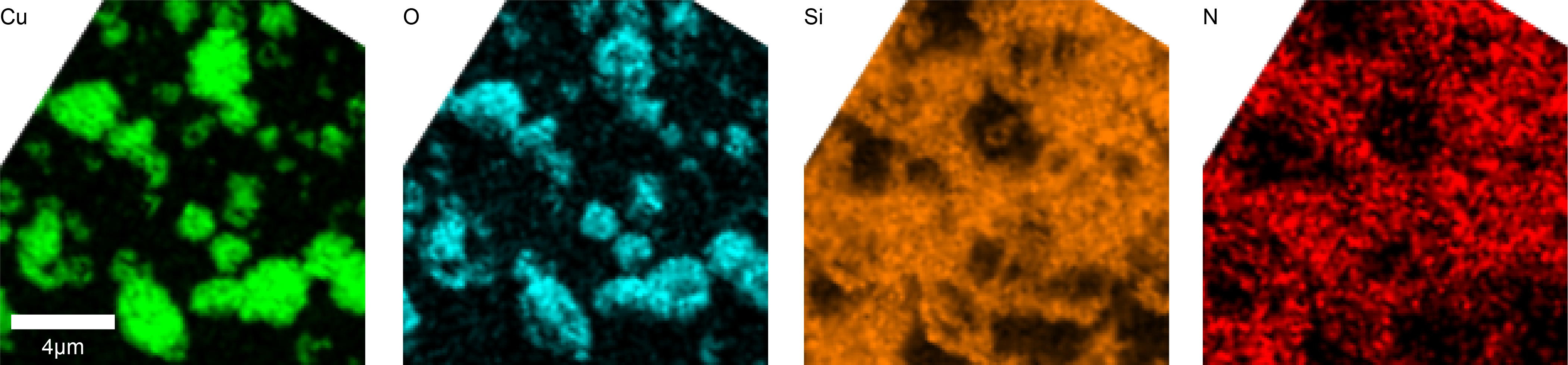}
\caption{\textbf{Post-mortem EDS analysis of ONEM pulsed electrodeposition.} The individual elemental EDS maps acquired in an SEM after the pulsed electrodeposition experiment shown in Fig.~\ref{fig:EC}i. The corresponding electron image and overlay map is shown in Fig.~\ref{fig:EC}h and i, respectively.}
\label{fig:EDX}
\end{figure}

\begin{figure}[H]
	\centering
	\includegraphics[width=1\textwidth]{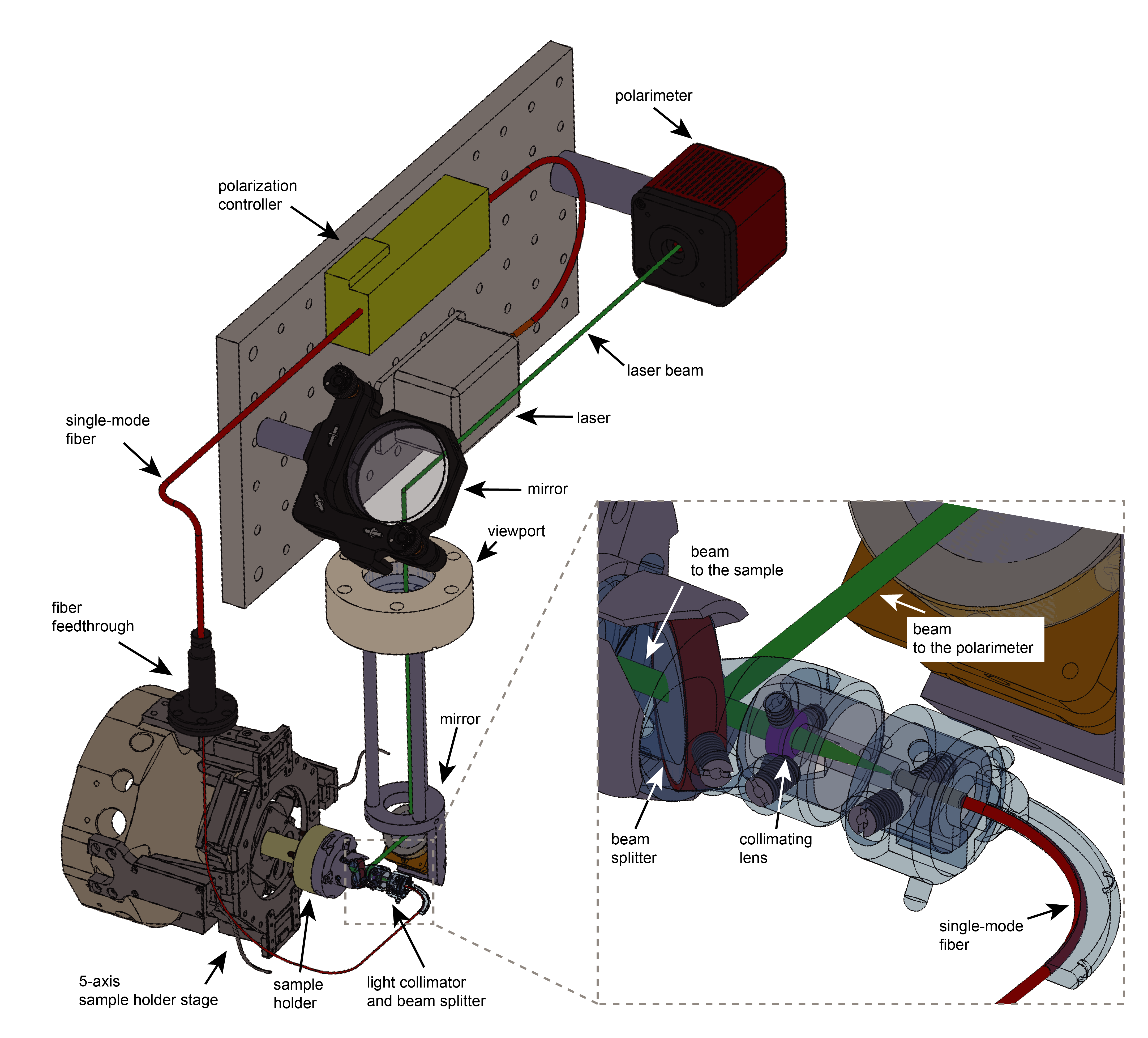}
	\caption{\textbf{Design of the optical module for the ONEM microscope.}
		Light is sent into the setup with a single-mode fiber. The collimator, followed by the 90/10 non-polarizing beam splitter, is mounted onto the sample holder and can be detached and stored inside the vacuum chamber during the sample transfer. The sample holder is mounted onto a 5-axis stage. The beam splitter transmits 90\% of the light to the sample and reflects 10\% to a polarimeter outside the vacuum chamber.}
	\label{fig:setup_solidworks} 
\end{figure}

\begin{figure}[H]
\centering
\includegraphics[width=0.5\textwidth]{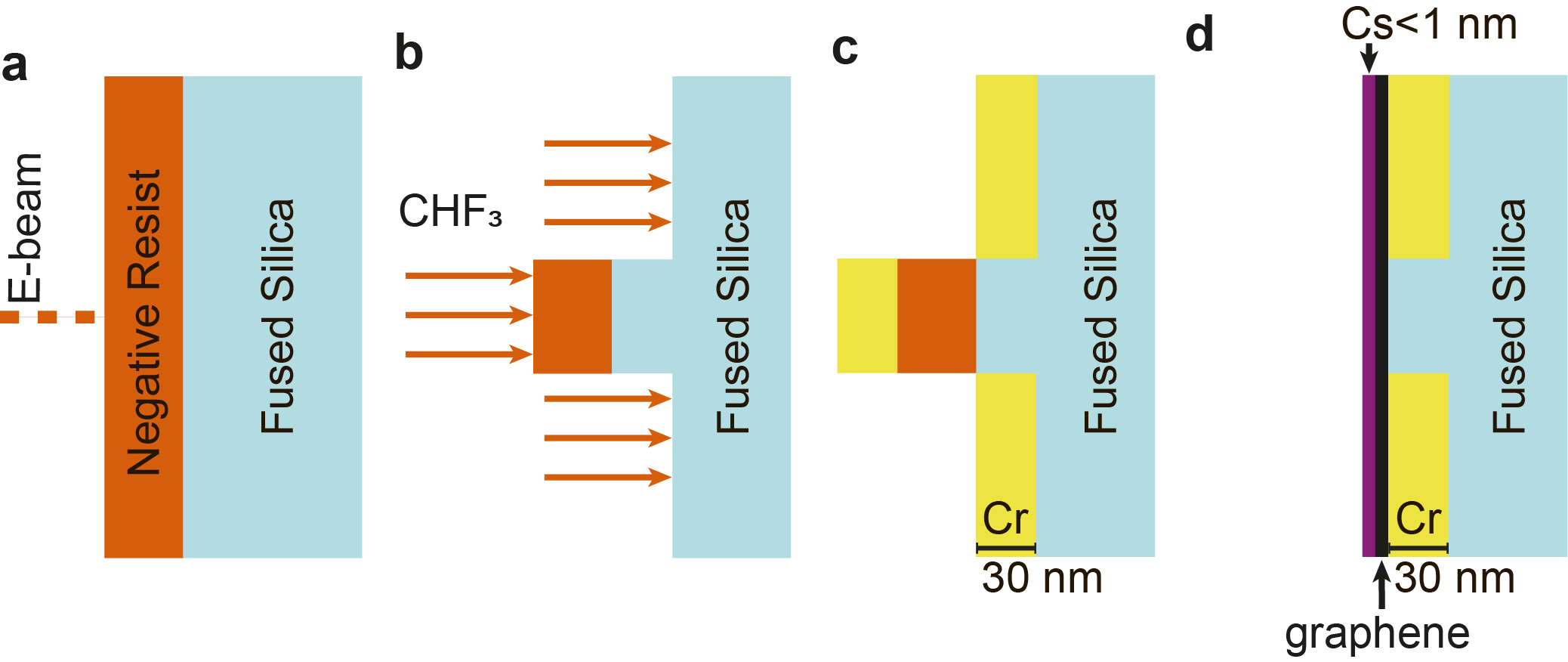}
\caption{\textbf{Sample preparation steps of the lithography sample.} Comprising, \textbf{a}, e-beam lithography, \textbf{b}, fused silica etching, \textbf{c}, chromium deposition, and \textbf{d}, photocathode deposition.}
\label{fig:lithography_prep}
\end{figure}

\begin{figure}[H]
\centering
\includegraphics[width=0.5\textwidth]{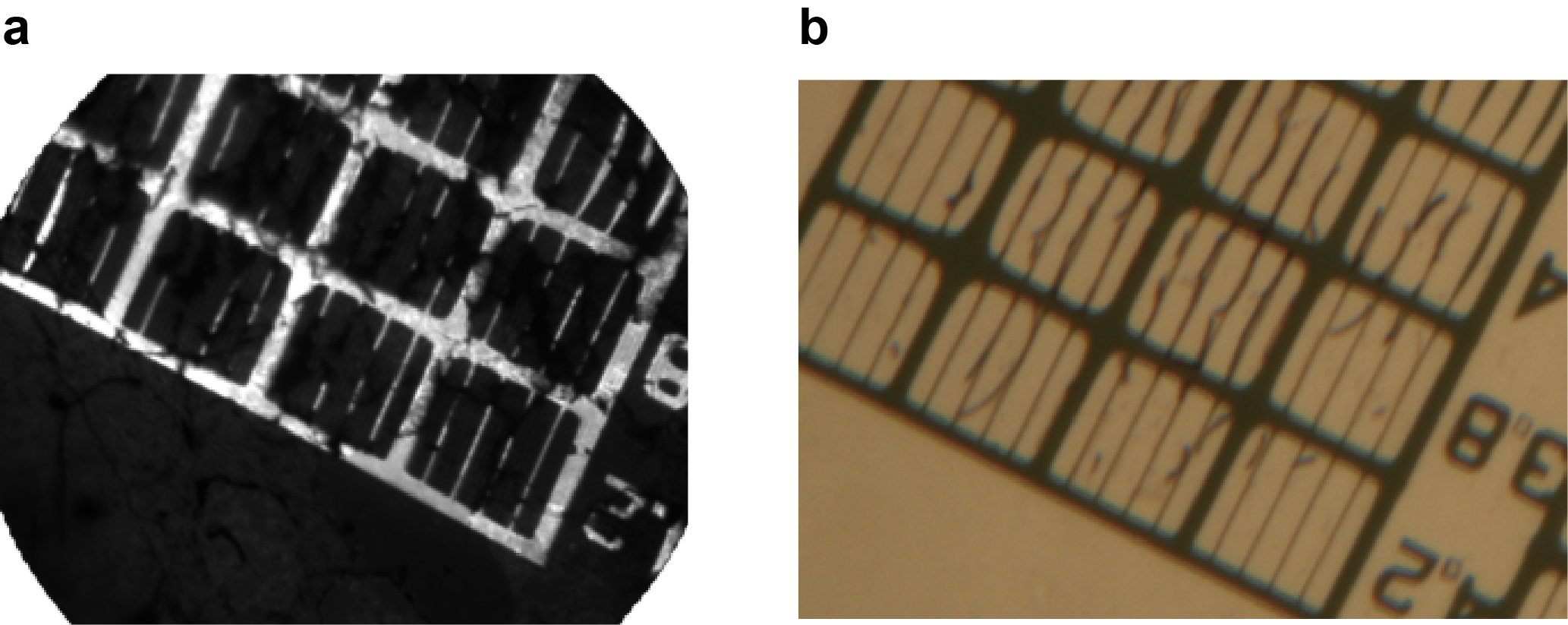}
\caption{\textbf{Images of the lithography sample.} \textbf{a}, ONEM image and \textbf{b} with an optical microscope. The squares have a side length of \SI{10}{\um} and may act as the scalebar.} 
\label{fig:lithography_intermediate_short}
\end{figure}

\begin{figure}[H]
\centering
\includegraphics[width=\textwidth]{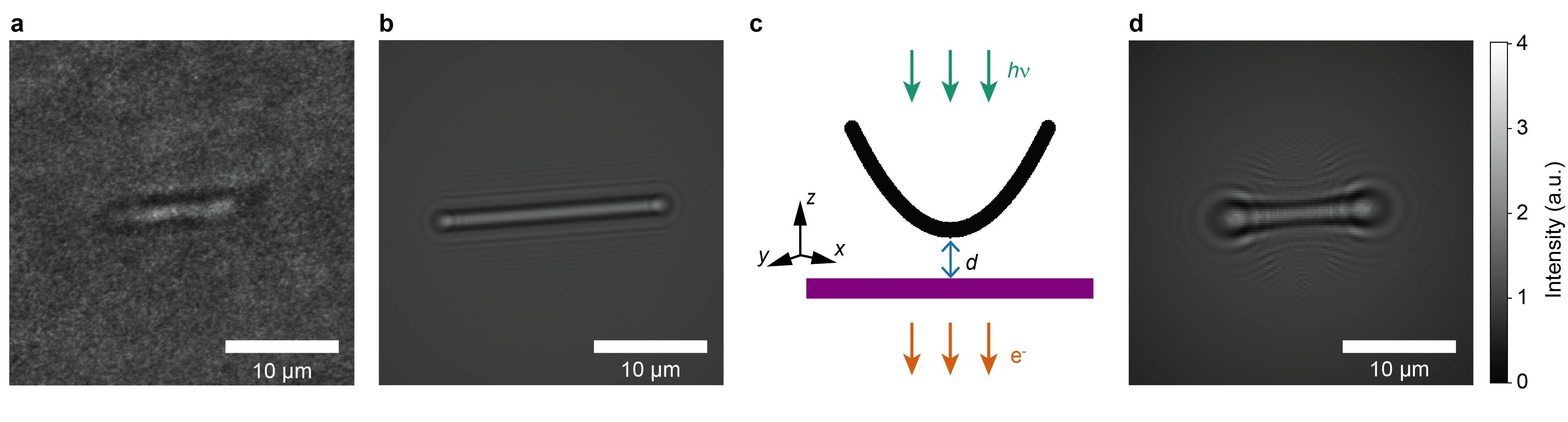}
\caption{\textbf{Bent \textit{E. coli} imaged with ONEM.} \textbf{a}, A frame of the data shown in the main text in Fig.~\ref{fig:ecoli}b (at $\SI{t = 7.3}{s}$). The bacterium appears shorter than in other frames. \textbf{b}, Simulated capsule-shaped bacterium with a length of \SI{20}{\um}, at a distance $d=\SI{3.15}{\um}$ from the interface, and at an angle of $\theta=0^\circ$ with respect to the interface. \textbf{c}, Sketch of a parabolic bacterium along with the incoming light in green, the outgoing photoelectrons in red, and the sample substrate and photocathode layer in purple. The distance to the substrate is again denoted as $d$.
\textbf{d}, Simulation result assuming a \SI{20}{\um} long parabolically shaped bacterium at $d=\SI{4}{\um}$.}
\label{fig:parabola}
\end{figure}

\begin{figure}[H]
\centering
\includegraphics[width=0.5\textwidth]{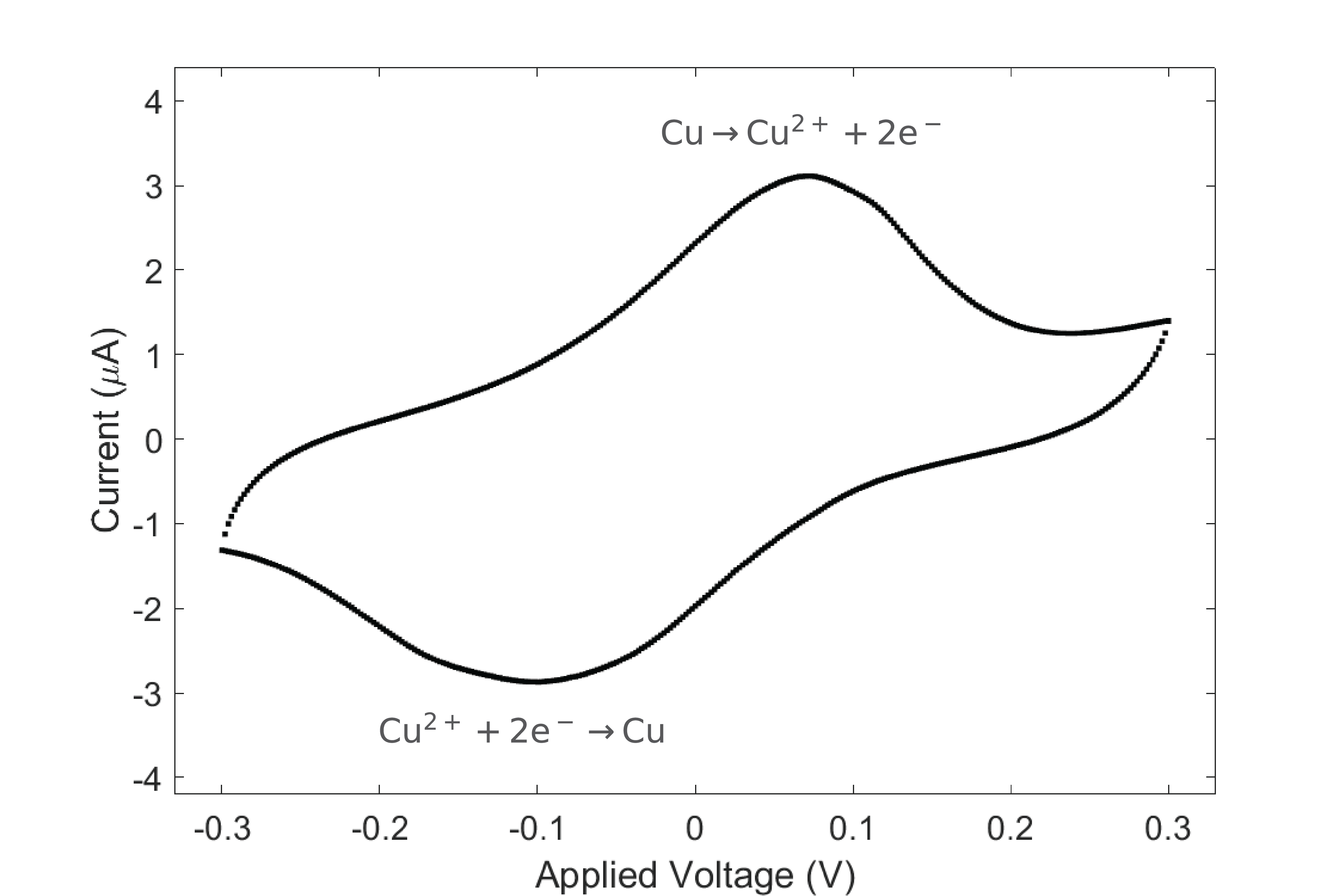}
\caption{\textbf{Table-top cyclic voltammetry in ONEM liquid cell.} Cyclic voltammogram recorded in the 2-electrode ONEM liquid cell for electrochemistry using a Gamry 600+ potentiostat in a table top experiment. The liquid cell was filled with $\mathrm{0.2 \, M \,CuSO_4 + 0.1\ M \,H_2SO_4}$ electrolyte. The potential sweep is carried out at the scan rate of \SI{100}{mV/s} from \SI{0.3}{V} to \SI{-0.3}{V} and reverse to complete the cycle.} 
\label{fig:gamry_cv}
\end{figure}

\begin{figure}[H]
\centering
\includegraphics[width=\textwidth]{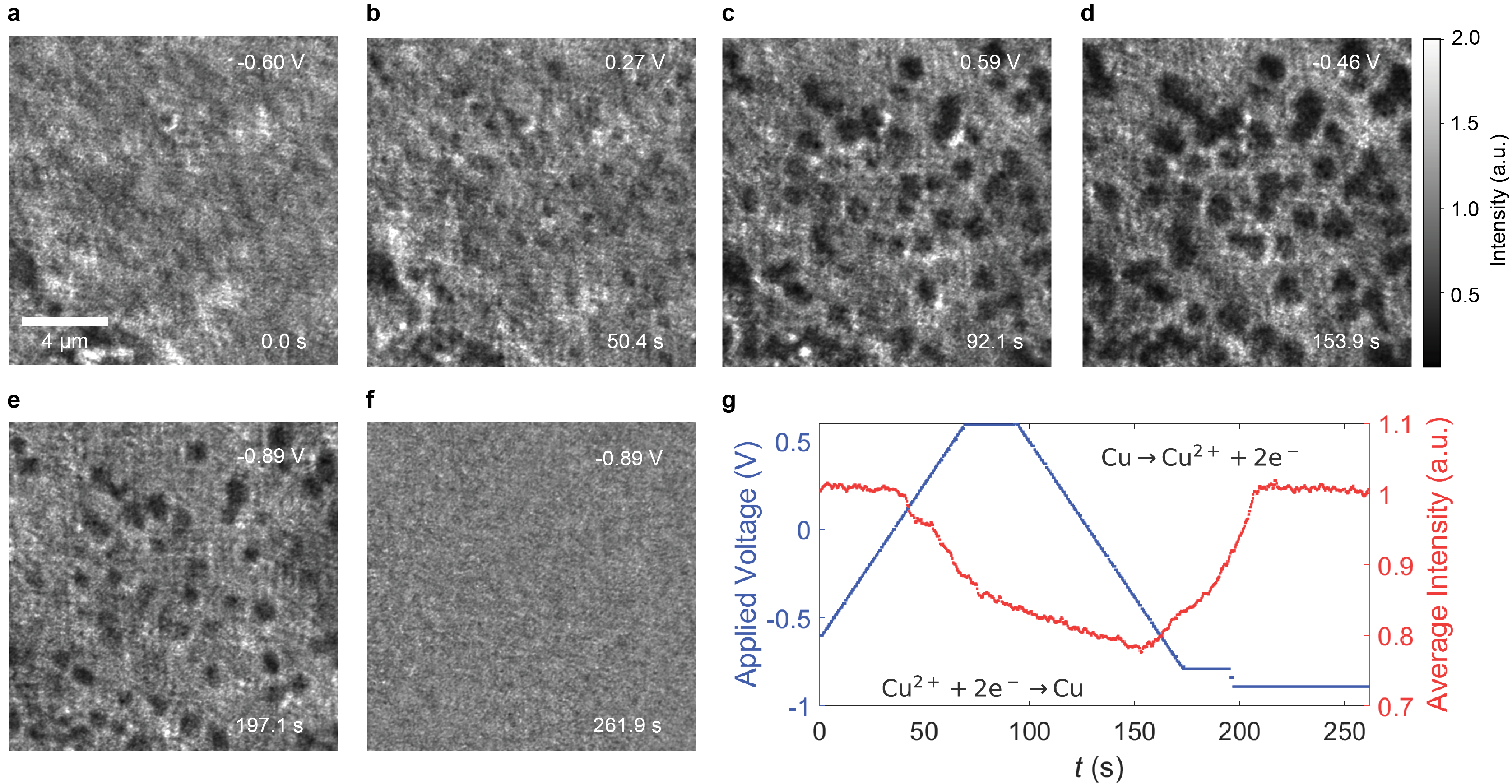}
\caption{\textbf{Cyclic potential sweep in ONEM liquid cell.} Cyclic potential sweep in a 2-electrode electrochemical ONEM liquid cell filled with $\mathrm{0.2 \, M \,CuSO_4 + 0.1\ M \,H_2SO_4}$ electrolyte. \textbf{a-f} shows sequential frames (Supplementary Video~5) acquired at different times while applying a linearly varying potential. \textbf{g} shows the applied potential in blue, and the average intensity of each frame in red. The intensity changes as a result of copper deposition and stripping. } 
\label{fig:cv}
\end{figure}

\newpage
\section*{Supplementary Information}
\subsection*{Supplementary Note 1: Relation between photoelectron flux and light intensity in the photocathode}
Our model assumes the photoelectron flux to be proportional to the intensity inside the photocathode.
This model is valid for light at normal incidence to the surface of the photocathode, with electric field vectors in the plane of the photocathode. For fields normal to the photocathode, the vectorial photoelectric effect has to be considered~\cite{kruger_attosecond_2018}. However, for the nanophotonic samples used in this work, the component of the field normal to the photocathode contributes to only 1\% of the total intensity. We can thus neglect this effect when comparing our data to FDTD simulations. 

\subsection*{Supplementary Note 2: Bending of the \textit{E. coli} in ONEM images} \label{sup_EC_LC}
Interestingly, in some frames bacteria appear to be significantly shorter than in most other images, for example at $\SI{t=7.3}{s}$, shown in Extended Data Fig.~\ref{fig:parabola}a, the simulation of a \SI{20}{\um} long capsule-shape bacterium does not match the data (see Extended Data Fig.~\ref{fig:parabola}b). Therefore, we speculate that the bacterium bends in the $z$-direction in these frames, resulting in a shorter appearance. Note that (sideways) bending of bacteria is also visible in other ONEM and optical microscopy videos, see Supplementary Video 3 and 2, respectively. To simulate a bent bacterium, we assume a parabolic shape, Extended Data Fig.~\ref{fig:parabola}c, resulting in an apparent length in accordance with the data. 

\subsection*{Supplementary Note 3: ONEM imaging of cyclic potential sweep in electrochemistry liquid cell} \label{sup_EC_LC} 
To demonstrate the potential of ONEM for electrochemistry experiments, we perform cyclic voltammetry (CV) in our setup, enabling the study of redox reactions~\cite{elgrishi_practical_2018, ross_liquid_2016}. 
Using the ONEM electrochemistry liquid cell described earlier, we carry out CV using a Gamry 600+ potentiostat.
In this table-top configuration, the working electrode (WE) and working sense (WS) terminals of the potentiostat were connected to the electrode on the \SiN chip, and the reference electrode (RE) and counter electrode (CE) terminals are connected to the electrode on the glass window. 
Extended Data Fig.~\ref{fig:gamry_cv} shows the resulting I-V curve obtained by cycling the potential between the two electrodes in an electrolyte consisting of $\mathrm{0.2 \, M \,CuSO_4 + 0.1\ M \,H_2SO_4}$ . The potential is swept linearly from \SI{0.3}{V} to \SI{-0.3}{V} and back, completing a cycle at a scan rate of \SI{100}{mV/s}. The cathodic peak corresponds to the reduction of $\mathrm{Cu^{2+}}$ to metallic copper on the WE ($\mathrm{Cu^{2+} + 2e^- \rightarrow Cu}$), while the anodic peak indicates the reverse oxidation process ($\mathrm{Cu \rightarrow Cu^{2+} + 2e^-}$).
This data, and its reproducibility, confirms proper electrical contact and electrode performance in the ONEM liquid cell. 

Next, we insert the ONEM electrochemical liquid cell into the ONEM microscope. Since the working electrode in our setup is tied to the sample stage of the electron microscope (at a potential of -10 \,kV), we vary the potential on the CE with respect to WE. As a result, the applied voltage is both rescaled and inverted relative to standard electrochemical conventions.
Still, it allows us to control and qualitatively understand the induced reactions.
Instead of measuring current, we monitor copper deposition in real-time using ONEM, which leads to a decrease in the average image intensity. Extended Data Fig.~\ref{fig:cv}g shows the average intensity and the applied potential as a function of time. 
As can be seen in Extended Data Fig.~\ref{fig:cv}g, the average intensity drops at a certain point during the linear increase of the applied potential, indicating the onset of copper deposition ($\mathrm{Cu^{2+}+e^-\rightarrow Cu}$). 
Extended Data Fig.~\ref{fig:cv}b shows this early stage of Cu island growth.
The potential is then held at a high positive value for approximately \SI{23}{s}, further increasing the amount of copper deposited (Extended Data Fig.~\ref{fig:cv}c). 
After starting the reverse potential sweep, copper deposition continued at a slower rate, until copper stripping ($\mathrm{Cu\rightarrow Cu^{2+}+e^-}$) is induced at a high negative potential.
The potential is further decreased and held at high negative values first leading to a decrease in cluster size (Extended Data Fig.~\ref{fig:cv}e) and subsequently to a clean WE substrate (Extended Data Fig.~\ref{fig:cv}f), completing the cyclic potential sweep experiment.
The data processing procedure is explained in the image processing section of the Methods section.

\subsection*{Supplementary Note 4: Pulsed electrodeposition data (Fig.~\ref{fig:EC})}
To verify the presence of deposited copper and to image it with an established technique, we use scanning electron microscopy (SEM). After acquiring the ONEM data, the liquid cell is opened and the \SiN chip is rinsed with deionized water. SEM measurements are performed 16 days after the ONEM experiment, during which the deposited copper has undergone morphological changes such as coarsening. The electron image and the corresponding overlaid EDS (energy-dispersive X-ray spectroscopy) images are shown in the main text in Fig.~\ref{fig:EC}h and i, respectively. Here, in Extended Data Fig.~\ref{fig:EDX} we present the individual elemental maps for the same field-of-view. To determine whether the crystallized structures could be attributed to salt residues from the electrolyte, we checked for the presence of sulfur. We did not detect any sulfur.



\clearpage 

\paragraph{Caption for Supplementary Video 1.}
\textbf{Live cell imaging using ONEM}
\\This video shows an ONEM live cell imaging of \textit{E. coli } in LB medium. Here we see a bacterium entering a static field of view, and its trajectory is overlaid in blue. Selected frames from this video, along with the trajectory of the bacterium, are presented in Fig.~\ref{fig:ecoli}. The video is recorded at a frame rate of \SI{3}{Hz}. The time stamps are displayed in the bottom-right corner.
\paragraph{Caption for Supplementary Video 2.}
\textbf{Live \textit{E. coli} after ONEM imaging, recorded with optical microscopy}
\\After ONEM live-cell imaging of the \textit{E. coli} bacteria, the liquid cell is examined with an optical microscope, imaging the liquid-facing side of the \SiN chip. In the video, the bacteria are seen to move around, similar to those in the ONEM videos. The continued presence of live bacteria confirms the non-invasiveness of the technique. This video is recorded using a $\mathrm{50\times}$ objective, at approximately 16 frames per second.
\paragraph{Caption for Supplementary Video 3.}
\textbf{Bacterium displaying curvature during motion}
\\To support the assumption that a bacterium may appear shorter due to its bending behavior (see Extended Data Fig.~\ref{fig:parabola}), Supplementary Video 3 is presented. While a large bacterium in the center adopts a non-linear geometry, we can see another bacterium enter the field of view and exhibit dynamic shape fluctuations. The videos are recorded at 3 Hz. Similar motion-related behavior is also apparent in the optical imaging of Supplementary Video 2.
\paragraph{Caption for Supplementary Video 4.}
\textbf{Electrochemical pulsed deposition}
\\This video presents electrochemical pulsed deposition over 11 cycles, imaged with ONEM (left), and synchronized with plots of the intensity standard deviation (red) versus applied potential (blue) on the right. Lower intensities indicate copper being deposited from the $\mathrm{0.2 \, M \,CuSO_4 + 0.1\ M \,H_2SO_4}$ electrolyte. The deposition occurs when a potential of \SI{0.5}{V} is applied. Representative frames from the video, along with the corresponding intensity standard deviation values for the entire experiment, are presented in Fig.~\ref{fig:EC}. The original video is acquired at \SI{3}{Hz}. Time values are shown in the bottom-right corner of each frame, and the applied voltage values are indicated in the top-right corner. 
\paragraph{Caption for Supplementary Video 5.}
\textbf{Cyclic potential sweep}
\\This video shows the deposition and stripping of Cu from $\mathrm{0.2 \, M \,CuSO_4 + 0.1\ M \,H_2SO_4}$ electrolyte during a cyclic potential sweep. The video on the left is synchronized to the plot of average intensity (red) of each frame versus the applied voltage (blue) on the right. Darker regions indicate copper deposition, with the clusters fading during oxidation. The time and applied potential values are in the bottom-right and top-right corners, respectively. The frame rate of the recorded video is \SI{3}{Hz}. Selected frames from the video are presented in Fig.~\ref{fig:cv}.

\end{document}